# Applications of Large Language Models in Radiation Oncology: From Workflow Automation to Clinical Intelligence

Yuzhen Ding, PhD[1], Jason Holmes, PhD[1], Yuexing Hao, PhD[1,2], Zhengliang Liu, PhD[3], Peilong Wang, PhD[4], Junjie Cui[5], Meiyun Cao, MS[1], Caiwen Jiang, PhD[1], Shuoyang Wei, PhD[1], Lin Zhao, PhD[6], Chenbin Liu, PhD[7], Lian Zhang, PhD[8,9,10,11], Yunze Yang, PhD[12], Tianming Liu, PhD[3], Wei Liu, PhD[1]

**Cc:** Yuzhen, Ding, Ph.D. <Ding.Yuzhen@mayo.edu>

Liu, Wei, Ph.D. <Liu.Wei@mayo.edu>

[1]Department of Radiation Oncology, Mayo Clinic, Phoenix, AZ 85054, USA

[2]Department of Electrical Engineering and Computer Science, Massachusetts Institute of Technology; Cambridge, MA, 02139, USA.

[3] School of Computing, University of Georgia, Athens, GA 30602, USA

[4] Department of radiation oncology, city of Hope national medical center, Duarte CA 91010,USA

[5] Department of Computer Science, Washington University in St. Louis, St. Louis, MO, 63130,USA

[6]Department of Biomedical Engineering, New Jersey Institute of Technology, Newark, NJ 07102, USA

[7]Department of Radiation Oncology, National Cancer Center/ National Clinical Research Center for Cancer / Cancer Hospital & Shenzhen Hospital, Chinese Academy of Medical Sciences and Peking Union Medical College, Shenzhen, 518116, China.

[8] Medical Artificial Intelligence Lab, The First Hospital of Hebei Medical University, Hebei Medical University, Shijiazhuang 050000, China.

[9]Hebei Provincial Medical Artificial Intelligence Research Institute, Shijiazhuang 050000, China.

[10]Hebei Provincial Engineering Research Center for AI-Based Cancer Treatment Decision-Making, The First Hospital of Hebei Medical University, Hebei Medical University, Shijiazhuang 050000, China.

[11]Department of Oncology, The First Hospital of Hebei Medical University, Hebei Medical University, Shijiazhuang 050000, China

[12]Department of Radiation Oncology, University of Miami, FL 33136, USA

## Abstract

Large language models (LLMs) have emerged as transformative tools in medicine, with strong capabilities in language understanding, reasoning, and structured information extraction. Radiation oncology is particularly well suited for LLM integration due to its data-intensive workflows, reliance on structured guidelines, and documentation burden. This review summarizes recent applications, including domain-specific fine-tuning for decision support, automated nomenclature standardization, registry curation using autonomous LLM agents, and protocol-aware radiotherapy plan evaluation using modular retrieval-augmented generation (RAG). Additional applications include patient safety analysis through incident classification and root cause analysis, electronic health record (EHR)-integrated communication, CT simulation order summarization, daily readiness briefings, and patient education systems. Emerging multimodal approaches enable context-aware contouring, while early studies show LLMs can assist treatment planning by interpreting dosimetric feedback. Together, these advances highlight a shift toward clinically grounded, auditable, and workflow-integrated AI systems that enhance efficiency, safety, and patient engagement.

## 1. Introduction

Radiation oncology is an information-dense specialty that combines diagnostic reasoning, protocol-driven decision making, treatment planning and delivery, longitudinal toxicity/outcomes tracking, and intensive patient communication[1-10]. While major advances, especially via the application of artificial intelligence in recent years, have been made in imaging[11-15], optimization-based planning[16], and delivery technologies[17], many day-to-day clinical and research workflows remain dominated by text: consultation notes, staging summaries, pathology and radiology reports, treatment prescriptions, simulation orders, plan evaluation narratives, and post-treatment follow-up documentation.

Large language models (LLMs) are a class of foundation models trained to predict and generate sequences of text. Modern LLMs are typically transformer-based neural networks pretrained on large corpora using self-supervised objectives[18]. Through this pretraining, LLMs learn statistical representations of language that can be adapted to downstream tasks such as summarization, question answering, classification, information extraction, and structured report generation. In clinical contexts, the utility of LLMs stems from their ability to interpret heterogeneous free-text inputs, track contextual dependencies across long documents, and translate unstructured narratives into structured outputs suitable for analytics[19-25]. Recently LLMs have been widely used in medicine due to their powerful capabilities[26-37].

Several technical capabilities are especially relevant to radiation oncology: (i) robust handling of jargon-rich medical language, (ii) the ability to follow complex instructions and output data in fixed formats (e.g., JSON), (iii) contextual reasoning across multiple sources (e.g., diagnoses, prior treatments, medications), and (iv) flexible interaction paradigms that enable conversational decision support for clinicians and educational assistance for patients.

However, deploying LLMs in high-stakes clinical settings requires careful attention to reliability and safety. A well-known challenge is hallucination—generation of plausible but incorrect statements—particularly when models are asked to answer questions without sufficient context or when source-of-truth data are unavailable. To mitigate this risk, modern clinical LLM systems increasingly rely on[38, 39]: (1) retrieval augmented generation (RAG)[40], where the model is grounded in explicit clinical data and guidelines; (2) fine-tuning LLMs using expert verified domain knowledge[41, 42]; and (3) modular architectures that separate deterministic components (e.g., constraint checking) from language generation[43].

From an implementation perspective, radiation oncology is well suited for such approaches because many critical decisions can be anchored to structured data (e.g., dose-volume statistics, fractionation schedules, modalities, and standard nomenclature) and because multiple consensus guidelines exist (e.g., NCCN, AAPM TG-263[44]). Furthermore, oncology information systems such as Aria, and enterprise electronic health record (EHR) systems such as Epic, provide structured anchors that can support auditable LLM deployments.

This review focuses on recent advances in applying LLMs to radiation oncology, with an emphasis on workflow-integrated, clinically relevant use cases. We first summarize foundational strategies for domain adaptation, including data curation, supervised fine-tuning, benchmarking, and grounded retrieval with modular system design. We then review applications across several domains: (a) clinical data integration and decision support, including treatment regimen generation, modality selection, and ICD-10 prediction; (b) standardization and quality improvement, including TG-263 structure renaming, registry curation, and protocol-aware radiotherapy plan evaluation; (c) automation of routine clinical tasks such as in-basket response message drafting, CT simulation order summarization, and daily readiness briefings for patient summarization and clinical trial matching; (d) emerging multimodal and reasoning-based applications in contouring, delineation, and treatment planning; and (e) patient-facing systems for education and engagement. Finally, we discuss key considerations for safety, validation, and implementation, and outline future directions for responsible integration into clinical practice. The overview of LLM applications in radiation oncology is shown in Figure 1.

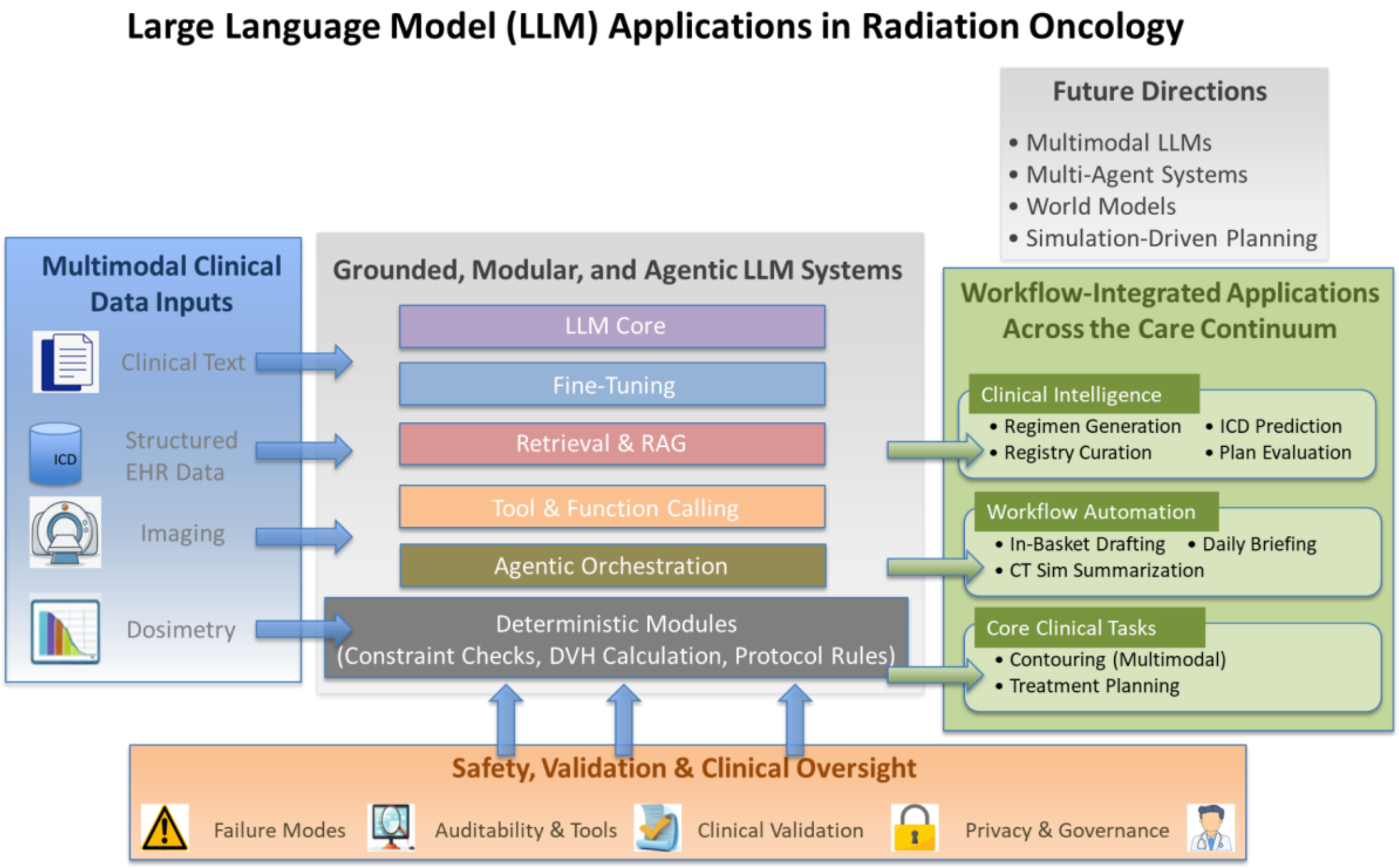

**Figure 1. Overview of LLM applications in radiation oncology.** Multimodal clinical data—including structured EHR data, free-text clinical notes, medical imaging, and dosimetric information—serve as inputs to grounded LLM systems. These systems incorporate domain adaptation (fine-tuning), RAG, tool use, and agentic orchestration to enable reliable and interpretable outputs. Applications span clinical data integration, workflow automation, core radiotherapy tasks (contouring and treatment planning), and patient-facing education. A cross-cutting safety and governance layer—including validation, auditability, and human oversight—is essential for clinical deployment. Emerging directions include multimodal LLMs, multi-agent systems, and world models for simulation-driven decision support.

## 2. Clinical Data Integration and Decision Support

To understand the role of LLMs in radiation oncology, it is important to first examine the foundational components that enable their effective deployment. These include domain-specific data curation, model adaptation through supervised fine-tuning, and systematic benchmarking of model performance in clinically relevant tasks. In addition, emerging frameworks emphasize structured integration with clinical data sources, guideline-driven standardization, and modular system design—such as RAG and tool-augmented agents—to improve reliability, interpretability, and auditability. The following sections review these core methodological and translational developments, highlighting how LLMs can be aligned with radiation oncology workflows while maintaining clinical validity and safety.

### 2.1 Domain-Specific Data Curation for LLM Training

High-performing domain models rely on high-quality training data. In radiation oncology, clinically meaningful supervision signals are often available, but they can be difficult to extract because relevant information is distributed across both structured fields and free-text clinical notes. To enable supervised fine-tuning, institutional patient data were retrieved using an in-house search engine that integrates structured data from the Varian ARIA database (version 15.6)[45]. Two datasets were assembled: (1) patient diagnosis data including patient identifiers, ICD codes, staging information, and free-text diagnosis details; and (2) treatment plan delivery details including course identifiers, plan types, modality, plan status, and fractionation parameters.

A total of 15,724 patient cases were initially extracted. To reduce ambiguity in supervision labels, cases with multiple diagnostic records or ambiguous primary plans were excluded, focusing instead on patients with a single diagnostic record and a clearly defined primary treatment plan. This filtering step reduced the dataset to 7,903 cases. Additional preprocessing improved consistency: planning information embedded within diagnosis notes (when present) was removed to avoid label leakage; punctuation was standardized; and initials were removed to improve generalization and privacy[45].

Diagnosis and plan datasets were then merged using standardized modality terms (e.g., photon, proton, electron, brachytherapy). When modality was not directly identified, manual annotation was performed based on original diagnostic data. ICD codes were already well documented, but only ICD-10 codes were retained, yielding a final set of 7,177 curated cases with precise ICD-10 labels for supervised fine-tuning of ICD prediction.

### 2.2 Supervised Fine-Tuning of Open-Source LLMs for Radiation Oncology Tasks

General-purpose LLMs often underperform in specialized clinical tasks due to domain shift, local practice patterns, and vocabulary differences. To address this, open-source models (LLaMA-2[46] 7B and Mistral[47] 7B) were fine-tuned using curated radiation oncology data for three tasks: treatment regimen generation, radiation modality selection, and ICD-10 code prediction[45]. Fine-tuning leveraged Low-Rank Adaptation (LoRA)[41], which freezes the pretrained model weights and injects trainable low-rank matrices into transformer layers. LoRA reduces computational cost and makes adaptation feasible on institutional infrastructure, enabling iterative experimentation while preserving base model capabilities.

For regimen generation and modality selection, 7,903 diagnosis-to-answer pairs were used (diagnostic summaries paired with clinician-annotated regimens and modality labels). For ICD-10 prediction, 7,177 diagnosis-to-code pairs were used. Model performance was evaluated using both automated metrics and clinician review to reflect real-world acceptability.

Across all tasks, fine-tuned models substantially outperformed their vanilla counterparts. For treatment regimen generation, the fine-tuned LLaMA-2 model achieved a ROUGE-1 score of 0.531 compared with 0.075 for the vanilla model, representing a statistically significant improvement ($p = 0.001$). For radiation modality selection, fine-tuned LLaMA-2 achieved 0.705 accuracy versus 0.499 for vanilla ($p = 0.001$). For ICD-10 code prediction, fine-tuned LLaMA-2 reached 0.642 accuracy versus 0.180 for vanilla ($p = 0.001$). Similar trends were observed for the fine-tuned Mistral model.

Beyond numerical metrics, clinical evaluation is critical because multiple reasonable regimens may exist for a given diagnosis depending on institutional practice patterns, patient comorbidities, and multidisciplinary recommendations. In clinicians' gradings, more than 60% of generated treatment regimens were considered clinically acceptable. These findings support the broader conclusion that open-source models can be aligned to radiation oncology practice via targeted fine-tuning and curation, providing a path toward cost-effective, customizable institutional assistants.

### 2.3 Benchmarking LLM Knowledge and Reasoning in Radiation Oncology Physics

In addition to adaptation for workflow tasks, understanding baseline domain competence is important for determining where LLMs can safely support clinicians and physicists. One study evaluated LLMs on a highly specialized topic: radiation oncology physics.[48] To minimize potential contamination (i.e., training data overlap), investigators constructed a novel 100-question multiple-choice examination designed to reflect the complexity of radiation oncology resident education. Two model versions were evaluated: GPT-3.5 and GPT-4.

GPT-4 demonstrated strong performance, achieving a higher average score than both GPT-3.5 and a cohort of practicing medical physicists. Interestingly, performance improved when the model was prompted to articulate its reasoning process, consistent with observations that structured prompting can enhance chain-of-reasoning quality and reduce shallow pattern-matching failures. Nevertheless, a collaborative team of medical physicists ultimately outperformed GPT-4, reinforcing that collective human expertise remains essential—particularly for ambiguous questions, contextual nuance, and safety-critical judgment.

From a translational standpoint, these benchmarking results suggest that LLMs can serve as useful assistants for knowledge recall, educational support, and preliminary reasoning—provided outputs are audited by domain experts. Such use may be particularly valuable for training, standard operating procedure development, and rapid reference tasks in physics and dosimetry.

Wang et al.[49] further evaluated the performance of recently released LLMs in answering radiation oncology physics questions. Our previously created 100 multiple-choice questions were reused, but with answer options randomly shuffled five times to have five new sets of exams. Five models—OpenAI o1-preview, GPT-4o, LLaMA 3.1 (405B), Gemini 1.5 Pro, and Claude 3.5 Sonnet (versions released before September 30, 2024)—were tested and compared with medical physicists' responses. Additional experiments replaced correct answers with "None of the above" to assess deductive reasoning and applied explain-first and step-by-step prompting to examine potential improvements. All models achieved expert-level performance, with o1-preview surpassing medical physicists by majority vote; however, performance dropped notably when correct options were replaced with "None of the above," suggesting there is still room for improvement on reasoning for LLaMA 3.1, Gemini 1.5 Pro, and Claude 3.5 Sonnet. Overall, the results demonstrate that modern LLMs can perform at an expert level on radiation oncology physics questions.

### 2.4 Standardizing Structure Nomenclature Using GPT-4 and TG-263

Structure naming standardization is a long-standing challenge in radiation oncology. Variability in region-of-interest (ROI) naming across planners, disease sites, and institutions reduces downstream usability of DICOM data for automation, analytics, and multi-center research. AAPM Task Group 263 provides a widely adopted nomenclature standard, but manual compliance remains burdensome.

To address this, a GPT-4–based system was benchmarked on its ability to relabel structure names according to TG-263[50]. The model was deployed as a DICOM storage server and provided with relevant excerpts from TG-263, including guiding principles and standardized structure name lists. A dataset of 600 patients spanning prostate, head and neck, and thorax cases was used for evaluation. Using this grounded approach, GPT-4 achieved high relabeling accuracies of 96.0%, 98.5%, and 96.9% for the respective disease sites.

These results suggest that LLMs can operationalize guideline language into consistent structured outputs, enabling rapid normalization of historical data. In practice, this may facilitate automated plan checking, atlas-based planning, dose accumulation studies, and outcomes research by reducing the manual preprocessing needed to unify nomenclature. As token limits expand and multimodal models mature, future systems may incorporate both text and geometric structure features to further improve robustness.

### 2.5 Autonomous LLM Agents for Outcomes Labeling and Registry Curation

High-quality outcomes registries are foundational for clinical research, quality improvement, and learning health systems. However, registry curation is often labor intensive and error prone due to reliance on heterogeneous documentation across structured and unstructured sources. A GPT-4o–based autonomous agent (RadOnc-GPT) was introduced to automate and audit outcomes labeling across clinical data[51].

The system was evaluated using a two-tier framework. In the first tier (quality assurance of retrieval), the agent achieved 100% accuracy in structured demographic retrieval and 99.4% accuracy in treatment plan extraction. In the second tier (generalization to outcomes labeling), the agent identified osteoradionecrosis (ORN) and cancer recurrence across prostate and head-and-neck cohorts (n = 895). After expert adjudication, accuracy improved to approximately 95–96%.

A key insight was that many discrepancies were attributable to institutional registry errors rather than model errors, with a majority of mismatches traced to ground-truth inaccuracies. Operationally, the agent executed function calls to retrieve EHR data, synthesized supporting evidence, and produced structured JSON summaries within approximately 10–30 seconds per patient. This design—modular, transparent, and self-retrieving—reduces reliance on free-form generation and supports auditability, making it attractive for continuous registry maintenance and real-time research cohort building.

### 2.6 Protocol-Aware Radiotherapy Plan Evaluation Using Modular RAG

Automated plan evaluation is an active area of research, but clinical adoption depends on reliability, traceability, and interpretability. A transparent and modular RAG framework was introduced for protocol-aware radiotherapy plan evaluation using a large foundation model (LLaMA-4 109B)[43]. The system integrates structured dose-volume data, population-based percentile scoring, and a clinical constraint-checking tool that references protocol-defined dose limits.

Using a curated dataset of 614 treatment plans across four disease sites, this approach employed a retrieval engine optimized via Gaussian Process tuning and a constraint module grounded in explicit protocol thresholds. The best-performing configuration in our experiments (with an all-MiniLM-L6-v2 retrieval backbone) achieved almost perfect nearest-neighbor accuracy within ±5 percentile points and less than 2 mean absolute error. End-to-end tests demonstrated 100% agreement between the LLM-generated summaries and independent module outputs, underscoring that the generative component was faithfully reporting deterministic calculations rather than inventing values.

This work highlights a design pattern likely to be critical for safe clinical deployment: separating deterministic logic (constraints, scoring, retrieval) from the narrative layer that communicates results. Such separation minimizes hallucination and supports explainability, while still benefiting from LLM fluency for producing human-readable plan assessment narratives.

### 2.7 Integrating LLMs for Patient Safety and Workflow Optimization in Radiation Oncology

Incident learning systems are essential for patient safety in radiation oncology, but extracting actionable insights from unstructured narratives remains resource-intensive. To address this bottleneck, approaches utilizing natural language processing and representational models have been developed; for instance, Beidler et al.[52] demonstrated transfer learning with representational models to automate the triaging of safety reports, while Zhang et al.[53] used statistical modeling for topic identification to uncover latent risks.

Transitioning to complex clinical reasoning, recent work leverages the advanced deductive capabilities of generative LLMs. Wang et al.[54] evaluated state-of-the-art generative models, including Gemini 2.5 Pro, for executing root cause analysis (RCA) on clinical incident reports, matching human expert analysis with over 90% accuracy and receiving high clinical validation. Expanding on this, a subsequent study[55] deployed an LLM to automatically convert 254 unstructured incidents into standardized taxonomies for quantitative risk assessment, successfully identifying critical failure pathways and high-risk predictors like procedural violations. As LLM performance benchmarks continue to mature, multi-agent systems will play a pivotal role in safety applications by autonomously coordinating clinical workflows, triaging incidents, notifying responsible parties, executing proactive Failure Mode and Effects Analysis (FMEA) and retrospective RCA, and driving continuous quality improvement.

## 3. Automation of Routine Clinical Tasks

Building on these foundational methods, LLMs are increasingly being integrated into routine clinical workflows and core tasks in radiation oncology. Many of these activities— ranging from patient communication and order interpretation to chart review, contouring, and treatment planning—require synthesis of heterogeneous data and iterative clinical reasoning, and are often time-intensive. LLM-based systems offer a potential approach to support these processes by generating structured summaries, assisting with decision-making, and augmenting domain-specific reasoning. The following sections highlight emerging applications in workflow automation as well as in clinically central tasks such as contouring and treatment planning, with an emphasis on workflow integration, efficiency gains, and the role of human oversight in ensuring safe deployment.

### 3.1 In-Basket Message Response Drafting with EHR-Integrated LLMs

Patient portal messaging has become a major channel for communication between patients and care teams. In radiation oncology, message content often spans appointments, treatment side effects, medication questions, PSA monitoring, imaging results, and insurance logistics. Post–COVID-19 increases in message volume have contributed to clinician burnout, and many messages are nonreimbursable yet time sensitive. Patients may also have variable health literacy, increasing the time required to craft clear and empathetic responses.

To address this challenge, an EHR-integrated system (RadOnc-GPT) was developed using GPT-4o to generate draft responses to in-basket messages[56]. The system can incorporate external context and patient-specific data retrieved from both enterprise EHR sources and radiation oncology–specific databases. In evaluation, 158 pre-recorded message interactions from 90 non-metastatic prostate cancer patients were analyzed across pre-treatment, on-treatment, and post-treatment phases.

Response quality was assessed using quantitative natural language processing (NLP) analysis and randomized single-blinded grading studies involving five clinicians and four nurses. Graders evaluated responses across domains including completeness, correctness, clarity, empathy, and estimated editing time. In clinician grading, RadOnc-GPT performed comparably to human care teams overall, with particularly strong performance in empathy and clarity, while human responses tended to score higher in completeness and correctness. Sentiment analysis (TextBlob and VADER) suggested that model outputs were more consistently positive in tone, whereas human responses spanned a broader sentiment range reflecting contextual nuance.

Operationally, clinician responders averaged approximately 3.6 minutes to draft responses and nurses averaged approximately 6.4 minutes, indicating substantial time investment. When used as a drafting tool, RadOnc-GPT can shift clinical work from writing to reviewing, potentially reducing response delays and workload. Limitations included occasional lack of clinical context, domain-specific inaccuracies, hallucinations, and inability to perform meta-tasks such as updating the medical record. These findings reinforce that safe deployment requires human oversight and guardrails, but also demonstrate that LLM drafting can yield meaningful efficiency gains.

### 3.2 CT Simulation Order Summarization with Locally Hosted LLMs

CT simulation orders contain key planning directives, but they are often entered as free text with variability in structure and completeness. Therapists commonly summarize physician orders for downstream workflow steps, creating opportunities for inconsistency and human error. Automating this summarization can improve standardization and reduce manual workload.

A locally hosted LLaMA 3.1 405B model was evaluated to automate summarization of physician-issued CT simulation orders[57]. A total of 607 orders were retrieved from Aria and categorized into seven groups by treatment modality and disease site. Customized prompts were co-developed with therapists to elicit consistent summaries. Ground-truth summaries were manually created and therapist verified, providing a reference standard for evaluation. The model achieved greater than 98% accuracy and produced summaries with improved consistency and readability compared with therapist-generated summaries. In addition, the approach significantly reduced the time required for summarization, highlighting its potential to improve workflow efficiency and reduce documentation burden. Importantly, local hosting can support institutional privacy requirements and enable integration into protected clinical environments. However, performance remains sensitive to variability and ambiguity in clinical documentation, and therapist oversight is still required for complex cases. Future direction may include exploring a broader range of treatment sites and further refining prompts to improve the robustness of the method. In addition, incorporating post-processing or validation checks may enhance reliability, particularly in cases with ambiguous or inconsistent inputs. Overall, this study provides a compelling demonstration of the potential for LLMs to streamline the CT simulation order summarization workflow and highlights that model performance remains closely tied to the clarity, structure, and internal consistency of the source documentation, especially in more complex clinical scenarios.

### 3.3 The Daily Dose: Morning Briefing for Patient Summary and Trial Matching

A persistent burden in radiation oncology is daily pre-clinic preparation: clinicians and care teams review long longitudinal records to understand disease status, prior therapies, imaging and pathology findings, toxicity trends, and pending decisions. In parallel, identifying clinical trial opportunities requires manual screening against complex eligibility criteria, which is often infeasible under routine time constraints. LLM-based clinical note summarization is an emerging approach to reduce this burden, with both general medicine and specialty applications demonstrating potential for chart review and discharge summary automation[58-60].

In The Daily Dose[61], a workflow-integrated pipeline generates a morning briefing email for participating radiation oncology physicians. The system is scheduled to run early each day, retrieves the clinician's schedule, and for each patient synthesizes a concise summary using structured EHR elements and recent clinical notes. Implementation details described in the group's draft include automated retries, institutional email delivery, and archiving/audit logging to support reliability and governance.

The briefing is designed to be "high signal" rather than exhaustive. Summaries emphasize diagnosis and stage, major oncologic events, treatment history (including radiation course details when available), and near-term action items tailored to visit type (e.g., consult vs. follow-up). This design choice aligns with broader findings in clinical summarization that usefulness depends strongly on information selection, structural consistency, and explicitly communicating uncertainty or missing data.

A distinctive component of The Daily Dose is automated clinical trial matching. For new or consulting visits, the system queries trial registries and uses LLM reasoning to screen patient-specific information against inclusion/exclusion criteria, producing a shortlist of candidate trials with brief rationales. This approach mirrors a rapidly developing literature on LLM-based trial matching (including TrialGPT-style approaches and multimodal pipelines), which reports clinically meaningful gains in screening accuracy and time-to-review compared with manual chart review[62-67].

In a post-deployment survey described in the draft (55 respondents out of 110 invited), clinicians reported favorable usability and satisfaction (mean 3.89±1.04 on a 5-point scale) and moderate-to-high perceived usefulness (mean 3.43±1.24). Respondents most consistently endorsed improved awareness of daily appointments and perceived help in identifying trial-eligible patients, and satisfaction was strongly associated with perceived time savings. Qualitative feedback highlighted improved preparedness but also emphasized the need to minimize inaccuracies and tailor content density to user preference.

These results suggest that "push" style LLM summaries (delivered proactively in an established channel like email) can be a practical complement to interactive copilots, provided that the system is tightly grounded in source data and incorporates ongoing quality monitoring. For future evaluations, reporting standards such as TRIPOD-LLM may help make performance claims more comparable across institutions and model versions[38].

### 3.4 LLM-Augmented Contouring and Delineation in Radiation Oncology

Contouring and delineation[68-71] are among the most labor-intensive and expertise-dependent tasks in radiation oncology. Although deep learning has improved auto-segmentation of organs at risk, target delineation remains more challenging because it depends not only on image appearance but also on clinical context, such as laterality, tumor stage, nodal status, and surgical history. This makes radiotherapy contouring a particularly relevant setting for LLM-augmented multimodal systems that combine imaging with text-rich clinical information.

A representative example is LLMSeg[72], an LLM-driven multimodal framework that integrates patient-specific clinical text with 3D imaging for context-aware target delineation. Rather than relying on image appearance alone, the model encodes key clinical information, such as tumor stage, surgery type, and laterality, and incorporates these language-derived features into the segmentation process to guide contour generation. In this way, LLMSeg is designed to better reflect the clinical reasoning underlying target definition, particularly when target boundaries cannot be reliably inferred from imaging alone. In breast cancer, LLMSeg achieved a Dice of 0.829 on the internal test set and maintained robust performance on two external test sets (0.822 and 0.844), whereas the vision-only baseline dropped to 0.731 and 0.444, respectively. Its advantage was also evident under limited training data, where the multimodal model maintained Dice above 0.8 with 40% of the training data and remained only slightly below 0.8 even with 20%, while the performance of the same framework using imaging input only deteriorated markedly and failed in the most data-limited setting.

These findings suggest that the primary value of LLMs in radiotherapy delineation lies not in achieving fully autonomous contouring, but in extending radiation oncology AI beyond vision-only segmentation by incorporating patient-specific clinical context into target definition, thereby enabling more clinically grounded delineation.

### 3.5 LLM-Guided Radiotherapy Treatment Planning

Radiotherapy treatment planning is a central responsibility of medical physicists and dosimetrists. In routine practice, plan optimization requires repeated adjustment of optimization objectives and priorities based on treatment planning system (TPS) feedback, particularly dose-volume histogram (DVH) metrics for target volumes and OARs. This iterative process is labor-intensive and experience-dependent. Because human-driven optimization is fundamentally a reasoning process that interprets dosimetric feedback and translates it into parameter adjustments, it is also a plausible target for LLM-based imitation.

Wei et al.[73] explored this concept in 35 single-target cervical cancer cases using two general-purpose LLMs: LLaMA-3.2, and Qwen-2.5-Max. The models were first provided with an explicit description of the reasoning workflow used by human physicists during iterative plan optimization. They were then exposed to five training cases through in-context learning, enabling them to infer optimization strategies from prior examples rather than through parameter fine-tuning. Under this framework, the LLMs were tasked with adjusting optimization settings across successive planning iterations on the basis of TPS-reported DVH indices.

Qwen-2.5-Max and LLaMA-3.2 demonstrated strong performance, completing automated optimization of a cervical cancer plan in a mean time of 16.3 and 9.8 minutes, respectively. After dose normalization, LLM-generated plans achieved significantly better conformity index and homogeneity index than manual plans ($p < 0.05$). For OAR sparing, plans produced by Qwen-2.5-Max and LLaMA-3.2 showed no significant differences from manual plans across most evaluated DVH indices. These findings suggest that LLMs may be capable of reproducing aspects of the physicist's iterative optimization workflow and provide preliminary evidence supporting their use in treatment planning assistance.

## 4. Patient-Facing LLM Systems for Education and Engagement

In addition to clinician-facing applications, LLMs are increasingly being explored for patient-facing use cases in radiation oncology. Patient education and engagement are essential components of cancer care, yet existing approaches are often limited by static content and variable accessibility. LLM-based systems offer a potential approach to address these limitations by generating explanations that are adaptive, interactive, and tailored to individual clinical contexts. By translating complex medical information into more accessible language while incorporating patient-specific data, these systems may complement traditional educational resources and support patient understanding throughout the care pathway.

### 4.1 Motivation: Information Gaps and Health Literacy in Oncology

A diagnosis of cancer frequently triggers substantial information seeking by patients and caregivers. However, publicly available resources vary in accuracy and level, and patients may encounter misinformation or become overwhelmed by technical terminology. In prostate cancer and other common radiotherapy indications, patients often seek clarification on staging, treatment options, expected side effects, timelines, and follow-up monitoring. Traditional educational materials are typically static and cannot easily adapt to the patient's specific clinical context.

### 4.2 MedEduChat: EHR-Integrated Closed-Domain Educational Agent

MedEduChat[74] was developed as a closed-domain LLM agent integrated with institutional EHR systems (Epic and Aria) to enhance prostate cancer patient education. Rather than relying solely on general web knowledge, the system retrieves patient-specific clinical data—such as diagnosis details, treatment history, clinical notes, and diagnostics—and generates layperson-friendly explanations aligned with NCCN guideline concepts. This retrieval-function-based approach aims to ground responses in source-of-truth clinical data and reduce hallucination risk.

In a mixed-method usability study involving 15 non-metastatic prostate cancer patients and three clinicians, MedEduChat achieved high usability (Usability Metric for User Experience (UMUX) ≈ 83.7/100) and significantly improved patient Health Confidence Scores (mean increase of approximately 4 points, $p < 0.05$). Clinician evaluations rated responses as highly correct (~2.9/3), complete (~2.7/3), safe (~2.7/3), and moderately personalized (~2.3/3). Qualitative interviews identified themes including patient 'unlearning and relearning' of misconceptions, the importance of trust, and the value of interactive explanations tailored to individual care trajectories.

Challenges included incomplete EHR data, emotional sensitivity for certain concerns, and off-scope patient queries that extend beyond the system's intended educational domain. These observations motivate continued development of scope control, escalation pathways, and clinician oversight mechanisms. Nevertheless, results suggest that EHR-integrated educational agents can improve patient engagement and confidence while complementing clinical care.

## 5. Safety, Validation, and Implementation Considerations

As LLM applications in radiation oncology expand from experimental use to clinical workflows, considerations of safety, validation, and implementation become increasingly central. While prior sections highlight promising use cases across data integration, workflow automation, and clinical decision support, real-world deployment requires systematic approaches to manage risk, ensure reliability, and maintain alignment with clinical standards. This includes identifying common failure modes, establishing grounding and auditability mechanisms, addressing data quality limitations, and defining appropriate evaluation and governance frameworks. The following sections summarize key considerations for translating LLM systems into safe and effective clinical practice.

### 5.1 Common Failure Modes in Clinical LLM Systems

Clinical deployment of LLMs requires explicit management of failure modes. Hallucination remains the most cited concern, but additional risks include: (i) omission of key clinical details (incompleteness), (ii) inappropriate generalization across disease sites or patient scenarios, (iii) overconfident tone when uncertainty should be communicated, (iv) prompt sensitivity and output variability, and (v) susceptibility to prompt injection or adversarial inputs when systems accept external text[38, 39].

In radiation oncology, such failures can manifest as incorrect fractionation recommendations, misinterpretation of prior therapy, omission of critical contraindications, or patient-facing messages that unintentionally minimize serious symptoms. Therefore, LLM outputs should generally be framed as draft suggestions or summaries, not autonomous clinical decisions, unless thoroughly validated and constrained[38].

### 5.2 Grounding and Auditability: RAG, Tool Use, and Modular Design

A consistent theme across successful applications is grounding. RAG systems reduce hallucination by injecting relevant evidence (e.g., guideline excerpts, structured EHR values, or protocol constraints) into the model's context. Tool-augmented agents go further by executing constrained function calls—such as querying plan databases or retrieving specific clinical fields—before generating a response. Modular architectures, as demonstrated in protocol-aware plan evaluation, isolate deterministic computations from narrative generation, enabling verification that reported values match source calculations[40, 43].

For outcomes labeling and registry curation, structured JSON outputs and explicit evidence trails improve transparency and facilitate human audit. For patient communication, grounding messages in retrieved clinical context helps ensure correctness, while templated safety checks (e.g., symptom escalation triggers) can reduce risk[51, 75].

### 5.3 Data Quality, Missingness, and Fact Verification

Clinical data quality is a limiting factor for downstream LLM performance. Missing values in structured variables and inconsistencies across documentation can compromise inference and retrieval. Two mitigation approaches were highlighted: (1) integrating fact-checking agents within RAG pipelines to verify outputs against structured sources and ensure reproducibility, and (2) applying deep denoising autoencoder (DAE)[76]–based imputation frameworks to reconstruct incomplete clinical variables. DAEs can learn latent representations that enable reconstruction of missing or noisy samples, potentially improving robustness of clinical factor extraction.

### 5.4 Evaluation: Beyond Automated Metrics

Evaluation must reflect clinical utility. Automated scores such as ROUGE or classification accuracy are useful but insufficient because multiple acceptable answers may exist. For clinical writing tasks (e.g., in-basket messaging), blinded clinician and nurse grading provides direct assessment of correctness, completeness, clarity, and empathy. For decision-support tasks, clinical acceptability grading and error taxonomy are critical. For standardization tasks (e.g., TG-263 renaming), accuracy can be directly measured against a gold standard, but robustness across disease sites and unusual structures should also be assessed[45, 50, 56].

Operational metrics—time saved, workload reduction, turnaround time, and user satisfaction—are essential for adoption. In addition, monitoring for drift and periodic revalidation are important as clinical practice patterns, protocols, and model versions evolve[38].

### 5.5 Privacy, Governance, and Deployment Considerations

Healthcare deployment requires compliance with privacy regulations and institutional governance. Key decisions include whether models are hosted locally versus accessed via secure cloud services, how PHI is handled and logged, and what safeguards exist for access control. Locally hosted models (e.g., for CT simulation order summarization) can reduce data exposure, whereas cloud-based models may provide stronger capabilities but require robust contractual and technical controls. Regardless of hosting, systems should incorporate least-privilege retrieval, audit logging, and clear escalation pathways when uncertainty or safety concerns arise[57, 75].

From a human factors standpoint, LLM systems should be designed to support clinician trust: outputs should cite evidence when possible, clearly express uncertainty, and be easy to edit. For patient-facing tools, scope limits and safety messaging are critical, especially for symptoms that require urgent evaluation[39, 74, 75].

## 6. Summary and Conclusions

LLM applications in radiation oncology are rapidly evolving from experimental demonstrations to workflow-integrated systems. The reviewed studies collectively indicate that clinically useful performance often depends on domain grounding—through fine-tuning on curated institutional data, retrieval of source-of-truth EHR fields, and modular architectures that constrain generation. Representative applications are summarized in Table 1. Within clinical data integration, fine-tuned open-source models improved regimen generation, modality selection, and ICD-10 prediction, while benchmarking studies established baseline competence in radiation oncology physics tasks. For standardization, GPT-4 achieved high accuracy in TG-263 structure relabeling, enabling data harmonization at scale. For clinical intelligence and quality improvement, autonomous agents demonstrated strong performance in outcomes labeling and registry auditing, while modular RAG systems enabled protocol-aware plan evaluation with traceability[43, 45, 50, 51]. Beyond these areas, LLMs have also been applied to workflow optimization—including in-basket message drafting, CT simulation order summarization, and daily readiness briefings—as well as to patient-facing education systems, multimodal contouring, and LLM-guided treatment planning, highlighting their expanding role across the radiation oncology care continuum.

A particularly important design pattern emerging across applications is the use of modular, protocol-aware systems that separate deterministic computation from generative reporting. In radiotherapy plan evaluation, retrieval-augmented frameworks integrating structured dose-volume data, population-based scoring, and explicit protocol constraints achieved high accuracy while maintaining full traceability between inputs and outputs[39]. By ensuring that LLM-generated narratives faithfully reflect independently computed metrics, such architectures reduce hallucination risk and enhance interpretability, addressing key barriers to clinical adoption.

In parallel, LLMs are increasingly being applied to patient safety and workflow optimization. Building on prior NLP-based triaging and topic modeling approaches[52, 53]. Recent generative LLM systems have demonstrated strong performance in higher-level reasoning tasks such as root cause analysis and incident classification[54, 55]. These models can transform unstructured safety reports into standardized taxonomies, identify latent failure pathways, and support proactive risk assessment. Looking forward, multi-agent LLM systems may further enable automated coordination of safety workflows, including incident triage, notification, Failure Mode and Effects Analysis (FMEA), and continuous quality improvement.

In operational workflows, EHR-integrated systems demonstrated the ability to draft in-basket message responses with strong clarity and empathy while reducing drafting time, and locally hosted models successfully summarized CT simulation orders with high accuracy and improved consistency. More recently, daily readiness briefings have been explored to streamline pre-clinic preparation by automatically summarizing scheduled patients and surfacing potential clinical trial opportunities within a department's active portfolio. On the patient-facing side, EHR-integrated educational agents improved patient health confidence and achieved high usability, supporting a vision of personalized, interactive education that complements clinician counseling[56, 57, 61, 74].

Beyond text-centric applications, recent advances highlight the emerging role of LLM-augmented multimodal systems in core radiotherapy tasks. In contouring and delineation, LLM-integrated frameworks enable the incorporation of patient-specific clinical context—such as tumor stage, laterality, and surgical history—into target definition, addressing limitations of vision-only segmentation[72]. These approaches demonstrate improved robustness, particularly in data-limited settings, and suggest a shift toward clinically grounded, context-aware delineation rather than fully autonomous contouring.

Similarly, in treatment planning, early studies indicate that LLMs can emulate aspects of the physicist's iterative optimization process by interpreting DVH feedback and adjusting planning parameters accordingly[73]. Using in-context learning, general-purpose models have demonstrated the ability to generate plans with comparable OARs sparing and improved conformity and homogeneity metrics relative to manual planning, while substantially reducing optimization time. These findings support the feasibility of LLM-guided planning assistance as a complement to existing treatment planning workflows.

Looking further ahead, multimodal LLMs and agentic architectures open transformative possibilities for radiation oncology. Current multimodal LLMs can jointly process clinical text, medical images (CT, MRI, PET), and structured dosimetric data within a unified reasoning framework, enabling capabilities that no single-modality system can achieve[77-79]. When extended to agentic architectures, such systems could coordinate specialized sub-agents for distinct tasks (e.g., image analysis, guideline retrieval, dose optimization, and safety checking), enabling end-to-end automation of complex radiotherapy pipelines while maintaining modularity and auditability at each step[80, 81].

A complementary and equally promising direction is the integration of world models for radiation oncology. World models, learned internal representations that simulate how a system evolves over time in response to actions, could predict the downstream consequences of planning decisions before they are executed[82, 83]. In the context of radiotherapy, a physics-informed world model could, for instance, anticipate how adjusting beam angles or dose constraints would affect the resulting dose distribution and OAR sparing, effectively performing "mental simulation" of the treatment planning process. Such predictive capacity would allow the agent to explore and compare multiple planning strategies internally, select the most promising approach, and present it to the physicist for review—dramatically reducing the number of iterative optimization cycles required. Beyond planning, world models could support longitudinal outcome prediction by modeling tumor response dynamics, normal tissue complication trajectories, and adaptive replanning triggers, ultimately moving radiation oncology toward anticipatory, simulation-driven clinical decision-making.

Across these use cases, safe deployment requires careful validation, human-in-the-loop oversight, and robust grounding. With these safeguards, LLM systems have the potential to become foundational components of next-generation radiation oncology practice—reducing documentation burden, improving data integrity, enhancing treatment planning efficiency, strengthening patient safety, and enabling more personalized patient engagement[38].

**Table 1. Summary of Representative LLM Applications in Radiation Oncology (Selected Examples)**

| Application | LLM Approach | Input Data | Output / Downstream Use | Key Evaluation | Clinical/Operational Impact |
|---|---|---|---|---|---|
| Fine-tuning open-source LLMs for treatment regimen generation, modality selection, and ICD-10 prediction | LoRA fine-tuning of LLaMA2-7B and Mistral-7B on institutional data; task-specific prompts | Curated diagnosis and treatment plan records from the oncology information system | Generated regimen text; modality labels; ICD-10 codes for downstream structuring and analytics | ROUGE-1 and accuracy; clinical grading of regimen acceptability | Demonstrated feasibility of domain adaptation with measurable improvements over base models |
| LLM benchmarking on radiation oncology physics questions | GPT-family models evaluated on domain-specific Q&A; error characterization | Expert-curated physics question set | Model answers; comparison across model generations | Accuracy and qualitative error analysis | Established baseline for safety/competence in specialized medical physics knowledge |
| TG-263 structure relabeling | GPT-4 prompting embedded | DICOM RT structure | Standardized structure names for data | Per-patient and per-structure | High accuracy suggests practical |
| and nomenclature standardization | in a DICOM server to relabel structure sets to TG-263 | sets (prostate, head & neck, thorax cohorts) | pooling and downstream analytics | accuracy against TG-263 standard | pathway for scaling nomenclature standardization |
| Autonomous outcomes labeling and registry curation (RadOnc-GPT) | Tool-augmented LLM agent retrieves EHR evidence, iteratively reasons, and outputs structured labels | Structured fields plus unstructured notes across multi-site cohorts | Outcomes labels (e.g., recurrence, osteoradionecrosis) with evidence trail | Tiered validation (QA tier + complex clinical outcomes tier) | Enables large-scale, auditable outcomes curation for research and quality improvement |
| Using a pretrained LLM as a clinical text encoder and training a 3D CNN segmentation model from scratch for target volume contouring in radiation oncology | Multimodal framework (LLMSeg) that uses a pretrained language model to encode clinical text and integrates it with a 3D CNN segmentation network via feature fusion | Breast radiotherapy CT images and patient-specific clinical text (tumor stage, laterality, surgery type, etc.) | Dice performance on internal (~0.83) and external (~0.82–0.84) test sets, with consistent improvements over the vision-only baseline and enhanced robustness under reduced training data. | Dice on internal (0.829) and external (0.822/0.844) test sets; comparison with vision-only models; robustness under reduced training data | Demonstrated that incorporating clinical context improves generalization and data efficiency, enabling more clinically grounded delineation |
| Utilizing LLMs for radiotherapy treatment planning by In-Context Learning | task-specific prompts, Qwen-2.5-Max, LLaMA-3.2 | DVH parameters of PTV and OARs from TPS | Dose objectives of OARs for next optimization iteration | DVH parameters of PTV and OARs | Demonstrated feasibility of using LLMs for automated radiotherapy treatment planning. |
| Protocol-aware radiotherapy plan evaluation | Modular RAG system with deterministic dose/constraint checks plus narrative generation | Treatment plan data, DVH metrics, protocol criteria, guideline text | Compliance assessment and explanations with traceability | Agreement with clinician review; error taxonomy | Supports consistent QA and protocol adherence auditing |
| EHR-integrated in-basket message drafting | Closed-domain LLM agent grounded in institutional EHR + specialty database | Historical in-basket messages and relevant EHR context | Draft responses scored for completeness, correctness, clarity, empathy | Clinician/nurse grading and time-to-edit estimates | Potential time savings and workload reduction while maintaining response quality |
| CT simulation order summarization with locally hosted LLMs | Locally hosted LLaMA 3.1 405B with therapist-designed prompts | Free-text CT simulation orders from Aria | Standardized summaries to guide CT simulation preparation | Error rate comparison vs. manual summaries; therapist feedback | Reduced variability, improved documentation consistency, and fewer human errors |
| Daily readiness briefing (The Daily Dose): patient summary + clinical trial matching | Schedule-triggered, EHR-integrated LLM agent; summarization + LLM-assisted trial screening | Daily clinic schedule; structured EHR fields; recent notes; trial registry content | Morning email brief with concise patient summaries and candidate trials with rationale | Clinician survey on usability, perceived usefulness, and qualitative feedback | Improved preparedness and trial awareness; highlighted need for accuracy and customization |
| EHR-integrated patient education agent (MedEduChat) | Patient-facing LLM agent grounded in EHR and validated | Patient EHR profile + curated educational content | Personalized explanations and answers to patient questions | Patient usability (UMUX) and clinician grading for correctness/safety | Improved patient health confidence and high usability with clinician oversight |
| | educational materials | | | | |

**References:**

1. Liu C, Liu Z, Holmes J, Zhang L, Zhang L, Ding Y, Shu P, Wu Z, Dai H, Li Y, Shen D, Liu N, Li Q, Li X, Zhu D, Liu T, Liu W. Artificial general intelligence for radiation oncology. Meta-Radiology. 2023;1(3):100045. doi: https://doi.org/10.1016/j.metrad.2023.100045.
2. Huynh E, Hosny A, Guthier C, Bitterman DS, Petit SF, Haas-Kogan DA, Kann B, Aerts HJWL, Mak RH. Artificial intelligence in radiation oncology. Nature Reviews Clinical Oncology. 2020;17(12):771-81. doi: 10.1038/s41571-020-0417-8.
3. Schild SE, Rule WG, Ashman JB, Vora SA, Keole S, Anand A, Liu W, Bues M. Proton beam therapy for locally advanced lung cancer: A review. World J Clin Oncol. 2014;5(4):568-75. Epub 2014/10/11. doi: 10.5306/wjco.v5.i4.568. PubMed PMID: 25302161; PMCID: PMC4129522.
4. Zhang X, Liu W, Li Y, Li X, Quan M, Mohan R, Anand A, Sahoo N, Gillin M, Zhu XR. Parameterization of multiple Bragg curves for scanning proton beams using simultaneous fitting of multiple curves. Phys Med Biol. 2011;56(24):7725-35. Epub 2011/11/17. doi: 10.1088/0031-9155/56/24/003. PubMed PMID: 22085829.
5. Matney J, Park PC, Bluett J, Chen YP, Liu W, Court LE, Liao Z, Li H, Mohan R. Effects of respiratory motion on passively scattered proton therapy versus intensity modulated photon therapy for stage III lung cancer: are proton plans more sensitive to breathing motion? International journal of radiation oncology, biology, physics. 2013;87(3):576-82. Epub 2013/10/01. doi: 10.1016/j.ijrobp.2013.07.007. PubMed PMID: 24074932; PMCID: PMC3825187.
6. Chen J, Yang Y, Feng H, Liu C, Zhang L, Holmes JM, Liu Z, Lin H, Liu T, Simone CB, Lee NY, Frank SJ, Ma DJ, Patel SH, Liu W. Enabling clinical use of linear energy transfer in proton therapy for head and neck cancer – A review of implications for treatment planning and adverse events study. Vis Cancer Med. 2025;6:3.
7. Chen J, Yang Y, Liu C, Feng H, Holmes J, Zhang L, Lin H, Lee N, Frank S, Simone CB, Ma DJ, Patel SH, Liu W. Critical review of patient outcome study in head and neck cancer radiotherapy. Meta-Radiology. 2025:100151. doi: https://doi.org/10.1016/j.metrad.2025.100151.
8. Deng W, Yang Y, Liu C, Bues M, Mohan R, Wong WW, Foote RH, Patel SH, Liu W. A Critical Review of LET-Based Intensity-Modulated Proton Therapy Plan Evaluation and Optimization for Head and Neck Cancer Management. Int J Part Ther. 2021;8(1):36-49. Epub 2021/07/22. doi: 10.14338/IJPT-20-00049.1. PubMed PMID: 34285934; PMCID: PMC8270082.
9. Li H, Li Y, Zhang X, Li X, Liu W, Gillin MT, Zhu XR. Dynamically accumulated dose and 4D accumulated dose for moving tumors. Med Phys. 2012;39(12):7359-67. Epub 2012/12/13. doi: 10.1118/1.4766434. PubMed PMID: 23231285; PMCID: PMC3523466.
10. Shan J, Yang Y, Schild SE, Daniels TB, Wong WW, Fatyga M, Bues M, Sio TT, Liu W. Intensity-modulated proton therapy (IMPT) interplay effect evaluation of asymmetric breathing with simultaneous uncertainty considerations in patients with non-small cell lung cancer. Med Phys. 2020;47(11):5428-40. Epub 2020/09/24. doi: 10.1002/mp.14491. PubMed PMID: 32964474; PMCID: PMC7722083.
11. Balagopal A, Morgan H, Dohopolski M, Timmerman R, Shan J, Heitjan DF, Liu W, Nguyen D, Hannan R, Garant A, Desai N, Jiang S. PSA-Net: Deep learning–based physician style–aware segmentation network for postoperative prostate cancer clinical target volumes. Artificial Intelligence in Medicine. 2021;121:102195. doi: https://doi.org/10.1016/j.artmed.2021.102195.
12. Ding Y, Feng H, Yang Y, Holmes J, Liu Z, Liu D, Wong WW, Yu NY, Sio TT, Schild SE. Deep‐learning based fast and accurate 3D CT deformable image registration in lung cancer. Medical Physics. 2023.

13. Ding Y, Holmes JM, Feng H, Li B, McGee LA, Rwigema J-CM, Vora SA, Ma DJ, Foote RL, Patel SH. Accurate Patient Alignment without Unnecessary Imaging Dose via Synthesizing Patient-specific 3D CT Images from 2D kV Images. arXiv preprint arXiv:240519338. 2024.
14. Zhang L, Liu Z, Zhang L, Wu Z, Yu X, Holmes J, Feng H, Dai H, Li X, Li Q, Wong WW, Vora SA, Zhu D, Liu T, Liu W. Technical note: Generalizable and promptable artificial intelligence model to augment clinical delineation in radiation oncology. Medical Physics. 2024;51(3):2187-99. doi: https://doi.org/10.1002/mp.16965.
15. Zhao Y, Yuan C, Liang Y, Li Y, Li C, Zhao M, Hu J, Zhong N, Liu W, Liu C. Streamlining Thoracic Radiotherapy Quality assurance: One-Class Classification for Automated OAR Contour Assessment. Technol Cancer Res Treat. 2025;24:15330338251345895. Epub 20250522. doi: 10.1177/15330338251345895. PubMed PMID: 40400421; PMCID: PMC12099094.
16. Zhang L, Holmes JM, Liu Z, Vora SA, Sio TT, Vargas CE, Yu NY, Keole SR, Schild SE, Bues M, Li S, Liu T, Shen J, Wong WW, Liu W. Beam mask and sliding window-facilitated deep learning-based accurate and efficient dose prediction for pencil beam scanning proton therapy. Med Phys. 2023. Epub 20230925. doi: 10.1002/mp.16758. PubMed PMID: 37748037.
17. Feng H, Shan J, Vargas CE, Keole SR, Rwigema J-CM, Yu NY, Ding Y, Zhang L, Schild SE, Wong WW, Vora SA, Shen J, Liu W. Online Adaptive Proton Therapy Facilitated by Artificial Intelligence-based Auto Segmentation in Pencil Beam Scanning Proton Therapy: Prostate oAPT in PBSPT. International Journal of Radiation Oncology*Biology*Physics. 2024. doi: https://doi.org/10.1016/j.ijrobp.2024.09.032.
18. Vaswani A, Shazeer N, Parmar N, Uszkoreit J, Jones L, Gomez AN, Kaiser L, Polosukhin I, editors. Attention is All you Need. Neural Information Processing Systems; 2017.
19. Dai H, Liu Z, Liao W, Huang X, Cao Y, Wu Z, Zhao L, Xu S, Zeng F, Liu W, Liu N, Li S, Zhu D, Cai H, Sun L, Li Q, Shen D, Liu T, Li X. AugGPT: Leveraging ChatGPT for Text Data Augmentation. IEEE Transactions on Big Data. 2025;11(03):907-18. doi: 10.1109/TBDATA.2025.3536934.
20. Liao W, Liu Z, Dai H, Wu Z, Zhang Y, Huang X, Chen Y, Jiang X, Liu D, Zhu D, Li S, Liu W, Liu T, Li Q, Cai H, Li X. Mask-guided BERT for few-shot text classification. Neurocomputing. 2024;610:128576. doi: https://doi.org/10.1016/j.neucom.2024.128576.
21. Xiao Z, Chen Y, Yao J, Zhang L, Liu Z, Wu Z, Yu X, Pan Y, Zhao L, Ma C, Liu X, Liu W, Li X, Yuan Y, Shen D, Zhu D, Yao D, Liu T, Jiang X. Instruction-ViT: Multi-modal prompts for instruction learning in vision transformer. Information Fusion. 2023:102204. doi: https://doi.org/10.1016/j.inffus.2023.102204.
22. Fang L, Yu X, Cai J, Chen Y, Wu S, Liu Z, Yang Z, Lu H, Gong X, Liu Y, Ma T, Ruan W, Abbasi A, Zhang J, Wang T, Latif E, Liu W, Zhang W, Kolouri S, Zhai X, Zhu D, Zhong W, Liu T, Ma P. Knowledge distillation and dataset distillation of large language models: emerging trends, challenges, and future directions. Artificial Intelligence Review. 2025;59(1):17. doi: 10.1007/s10462-025-11423-3.
23. Ruan W, Lyu Y, Zhang J, Cai J, Shu P, Ge Y, Lu Y, Gao S, Wang Y, Wang P, Zhao L, Wang T, Liu Y, Fang L, Liu Z, Liu Z, Li Y, Wu Z, Chen J, Jiang H, Pan Y, Yang Z, Chen J, Liang S, Zhang W, Ma T, Dou Y, Zhang J, Gong X, Gan Q, Zou Y, Chen Z, Qian Y, Yu S, Lu J, Song K, Wang X, Sikora A, Li G, Li X, Li Q, Wang Y, Zhang L, Abate Y, He L, Zhong W, Liu R, Huang C, Liu W, Shen Y, Ma P, Zhu H, Yan Y, Zhu D, Liu T. Large language models for bioinformatics. Quantitative Biology. 2026;14(1):e70014. doi: https://doi.org/10.1002/qub2.70014.
24. Jiang H, Pan Y, Chen J, Liu Z, Sun L, Li Q, Zhang L, Zhu D, Wang X, Liu W, Li X, Li G, Zhang W, Zhao L, Yu X, Wang Y, Liu T. Quantum artificial intelligence: A comprehensive survey. Meta-Radiology. 2026;4(1):100205. doi: https://doi.org/10.1016/j.metrad.2026.100205.
25. Li X, Zhao L, Zhang L, Wu Z, Liu Z, Jiang H, Cao C, Xu S, Li Y, Dai H, Yuan Y, Liu J, Li G, Zhu D, Yan P, Li Q, Liu W, Liu T, Shen D. Artificial General Intelligence for Medical

Imaging Analysis. IEEE Reviews in Biomedical Engineering. 2024. doi: 10.1109/RBME.2024.3493775.
26. Liu Z, He M, Jiang Z, Wu Z, Dai H, Zhang L, Luo S, Han T, Li X, Jiang X, Zhu D, Cai X, Ge B, Liu W, Liu J, Shen D, Liu T. Survey on natural language processing in medical image analysis. Zhong Nan Da Xue Xue Bao Yi Xue Ban. 2022;47(8):981-93. Epub 2022/09/14. doi: 10.11817/j.issn.1672-7347.2022.220376. PubMed PMID: 36097765.
27. Liu Z, Li Y, Shu P, Zhong A, Jiang H, Pan Y, Yang L, Ju C, Wu Z, Ma C, Chen C, Kim S, Dai H, Zhao L, Sun L, Zhu D, Liu J, Liu W, Shen D, Li Q, Liu T, Li X. Radiology-GPT: A large language model for radiology. Meta-Radiology. 2025;3(2):100153. doi: https://doi.org/10.1016/j.metrad.2025.100153.
28. Liu Z, Xu S, Wu Z, Murray B, Barreto E, Most A, Li S, Liu W, Li X, Liu T, Sikora A. PharmacyGPT: Exploration of Artificial Intelligence for Medication Management in the Intensive Care Unit,. BMC Medical Informatics and Decision Making. 2025.
29. Liu Z, Zhang L, Wu Z, Yu X, Cao C, Dai H, Liu N, Liu J, Liu W, Li Q, Shen D, Li X, Zhu D, Liu T. Surviving ChatGPT in healthcare. Front Radiol. 2023;3:1224682. Epub 20240223. doi: 10.3389/fradi.2023.1224682. PubMed PMID: 38464946; PMCID: PMC10920216.
30. Liu Z, Zhong A, Li Y, Yang L, Ju C, Wu Z, Ma C, Shu P, Chen C, Kim S, Dai H, Zhao L, Zhu D, Liu J, Liu W, Shen D, Li Q, Liu T, Li X, editors. Tailoring Large Language Models to Radiology: A Preliminary Approach to LLM Adaptation for a Highly Specialized Domain. Machine Learning in Medical Imaging; 2024 2024//; Cham: Springer Nature Switzerland.
31. Rezayi S, Dai H, Liu Z, Wu Z, Hebbar A, Burns AH, Zhao L, Zhu D, Li Q, Liu W, Li S, Liu T, Li X, editors. ClinicalRadioBERT: Knowledge-Infused Few Shot Learning for Clinical Notes Named Entity Recognition. Machine Learning in Medical Imaging; 2022 2022//; Cham: Springer Nature Switzerland.
32. Zhang K, Zhou R, Adhikarla E, Yan Z, Liu Y, Yu J, Liu Z, Chen X, Davison BD, Ren H, Huang J, Chen C, Zhou Y, Fu S, Liu W, Liu T, Li X, Chen Y, He L, Zou J, Li Q, Liu H, Sun L. A generalist vision–language foundation model for diverse biomedical tasks. Nature Medicine. 2024. doi: 10.1038/s41591-024-03185-2.
33. Wu Z, Zhang L, Cao C, Yu X, Liu Z, Zhao L, Li Y, Dai H, Ma C, Li G, Liu W, Li Q, Shen D, Li X, Zhu D, Liu T. Exploring the Trade-Offs: Unified Large Language Models vs Local Fine-Tuned Models for Highly-Specific Radiology NLI Task. IEEE Transactions on Big Data. 2025;11(3):1027-41. doi: 10.1109/TBDATA.2025.3536928.
34. Zhong T, Zhao W, Zhang Y, Pan Y, Dong P, Jiang Z, Jiang H, Zhou Y, Kui X, Shang Y, Zhao L, Yang L, Wei Y, Li Z, Zhang J, Yang L, Chen H, Zhao H, Liu Y, Zhu N, Li Y, Wang Y, Yao J, Wang J, Zeng Y, He L, Zheng C, Zhang Z, Li M, Liu Z, Dai H, Wu Z, Zhang L, Zhang S, Cai X, Hu X, Zhao S, Jiang X, Zhang X, Liu W, Li X, Zhu D, Guo L, Shen D, Han J, Liu T, Liu J, Zhang T. ChatRadio-Valuer: A Chat Large Language Model for Generalizable Radiology Impression Generation on Multi-institution and Multi-system Data. IEEE Transactions on Biomedical Engineering. 2025:1-12. doi: 10.1109/TBME.2025.3597325.
35. Zhao Y, Zhong E, Yuan C, Li Y, Zhao M, Li C, Hu J, Liu W, Liu C. Med-VLM: Enhancing Medical Image Segmentation Accuracy through Vision-Language Model. 2025 ICCV PHAROS-AFE-AIMI Workshop 2025.
36. Liu Z, Zhong T, Li Y, Zhang Y, Pan Y, Zhao Z, Dong P, Cao C, Liu Y, Shu P, Wei Y, Wu Z, Ma C, Wang J, Wang S, Zhou M, Jiang Z, Li C, Holmes J, Xu S, Zhang L, Dai H, Zhang K, Zhao L, Chen Y, Liu X, Wang P, Yan P, Liu J, Ge B, Sun L, Zhu D, Li X, Liu W, Cai X, Hu X, Jiang X, Zhang S, Zhang X, Zhang T, Zhao S, Li Q, Zhu H, Shen D, Liu T. Evaluating Large Language Models for Radiology Natural Language Processing2023 July 01, 2023:[arXiv:2307.13693 p.]. Available from: https://ui.adsabs.harvard.edu/abs/2023arXiv230713693L.
37. Liu Z, Jiang H, Zhong T, Wu Z, Ma C, Li Y, Yu X, Zhang Y, Pan Y, Shu P, Lyu Y, Zhang L, Yao J, Dong P, Cao C, Xiao Z, Wang J, Zhao H, Xu S, Wei Y, Chen J, Dai H, Wang P,

He H, Wang Z, Wang X, Zhang X, Zhao L, Liu Y, Zhang K, Yan L, Sun L, Liu J, Qiang N, Ge B, Cai X, Zhao S, Hu X, Yuan Y, Li G, Zhang S, Zhang X, Jiang X, Zhang T, Shen D, Li Q, Liu W, Li X, Zhu D, Liu T. Holistic Evaluation of GPT-4V for Biomedical Imaging2023 November 01, 2023:[arXiv:2312.05256 p.]. Available from: https://ui.adsabs.harvard.edu/abs/2023arXiv231205256L.
38. Gallifant J, Afshar M, Ameen S, Aphinyanaphongs Y, Chen S, Cacciamani G, Demner-Fushman D, Dligach D, Daneshjou R, Fernandes C, Hansen LH, Landman A, Lehmann L, McCoy LG, Miller T, Moreno A, Munch N, Restrepo D, Savova G, Umeton R, Gichoya JW, Collins GS, Moons KGM, Celi LA, Bitterman DS. The TRIPOD-LLM reporting guideline for studies using large language models. Nature Medicine. 2025;31(1):60-9. doi: 10.1038/s41591-024-03425-5.
39. Pais C, Liu J, Voigt R, Gupta V, Wade E, Bayati M. Large language models for preventing medication direction errors in online pharmacies. Nature Medicine. 2024;30(6):1574-82. doi: 10.1038/s41591-024-02933-8.
40. Lewis P, Perez E, Piktus A, Petroni F, Karpukhin V, Goyal N, Kuttler H, Lewis M, Yih W-t, Rocktäschel T, Riedel S, Kiela D. Retrieval-Augmented Generation for Knowledge-Intensive NLP Tasks. ArXiv. 2020;abs/2005.11401.
41. Hu JE, Shen Y, Wallis P, Allen-Zhu Z, Li Y, Wang S, Chen W. LoRA: Low-Rank Adaptation of Large Language Models. ArXiv. 2021;abs/2106.09685.
42. Goswami J, Prajapati KK, Saha A, Saha AK. Parameter-efficient fine-tuning large language model approach for hospital discharge paper summarization. Applied Soft Computing. 2024;157:111531. doi: https://doi.org/10.1016/j.asoc.2024.111531.
43. Cui J, Wang P, Holmes JM, Sun L, Hinni ML, Pockaj BA, Vora SA, Sio TT, Wong WW, Yu NY, Schild SE, Niska JR, Keole SR, Rwigema J-CM, Patel SH, McGee LA, Vargas CA, Liu W. An Automated Retrieval-Augmented Generation LLaMA-4 109B-based System for Evaluating Radiotherapy Treatment Plans. ArXiv. 2025;abs/2509.20707.
44. Mayo CS, Moran JM, Bosch W, Xiao Y, McNutt T, Popple R, Michalski J, Feng M, Marks LB, Fuller CD, Yorke E, Palta J, Gabriel PE, Molineu A, Matuszak MM, Covington E, Masi K, Richardson SL, Ritter T, Morgas T, Flampouri S, Santanam L, Moore JA, Purdie TG, Miller RC, Hurkmans C, Adams J, Jackie Wu QR, Fox CJ, Siochi RA, Brown NL, Verbakel W, Archambault Y, Chmura SJ, Dekker AL, Eagle DG, Fitzgerald TJ, Hong T, Kapoor R, Lansing B, Jolly S, Napolitano ME, Percy J, Rose MS, Siddiqui S, Schadt C, Simon WE, Straube WL, St James ST, Ulin K, Yom SS, Yock TI. American Association of Physicists in Medicine Task Group 263: Standardizing Nomenclatures in Radiation Oncology. International journal of radiation oncology, biology, physics. 2018;100(4):1057-66. Epub 20171215. doi: 10.1016/j.ijrobp.2017.12.013. PubMed PMID: 29485047; PMCID: PMC7437157.
45. Wang P, Liu Z, Li Y, Holmes J, Shu P, Zhang L, Li X, Li Q, Laughlin BS, Toesca DS, Vargas CE, Vora SA, Patel SH, Sio TT, Liu T, Liu W. Fine-tuning open-source large language models to improve their performance on radiation oncology tasks: A feasibility study to investigate their potential clinical applications in radiation oncology. Medical Physics. 2025;52(7):e17985. doi: https://doi.org/10.1002/mp.17985.
46. Touvron H, Lavril T, Izacard G, Martinet X, Lachaux M-A, Lacroix T, Rozière B, Goyal N, Hambro E, Azhar F, Rodriguez Ae, Joulin A, Grave E, Lample G. LLaMA: Open and Efficient Foundation Language Models. ArXiv. 2023;abs/2302.13971.
47. Jiang AQ, Sablayrolles A, Mensch A, Bamford C, Chaplot DS, Casas DdL, Bressand F, Lengyel G, Lample G, Saulnier L, Lavaud LeR, Lachaux M-A, Stock P, Scao TL, Lavril T, Wang T, Lacroix T, Sayed WE. Mistral 7B. ArXiv. 2023;abs/2310.06825.
48. Holmes J, Liu Z, Zhang L, Ding Y, Sio TT, McGee LA, Ashman JB, Li X, Liu T, Shen J, Liu W. Evaluating large language models on a highly-specialized topic, radiation oncology physics. Front Oncol. 2023;13. doi: 10.3389/fonc.2023.1219326.

49. Wang P, Holmes J, Liu Z, Chen D, Liu T, Shen J, Liu W. A recent evaluation on the performance of LLMs on radiation oncology physics using questions of randomly shuffled options. Front Oncol. 2025;Volume 15 - 2025. doi: 10.3389/fonc.2025.1557064.
50. Holmes J, Zhang L, Ding Y, Feng H, Liu Z, Liu T, Wong WW, Vora SA, Ashman JB, Liu W. Benchmarking a Foundation Large Language Model on its Ability to Relabel Structure Names in Accordance With the American Association of Physicists in Medicine Task Group-263 Report. Pract Radiat Oncol. 2024. Epub 20240726. doi: 10.1016/j.prro.2024.04.017. PubMed PMID: 39243241.
51. Holmes JM, Hao Y, Borras-Osorio M, Mastroleo F, Romero-Brufau S, Carducci V, Abel KMV, Routman DM, Foong AYK, Muller LM, Shiraishi S, Ebner DK, Ma DJ, Keole SR, Patel SH, Fatyga M, Bues M, Stish BJ, Garces YI, Wittich MAN, Foote RL, Vora SA, Laack NN, Waddle MR, Liu W. RadOnc-GPT: An Autonomous LLM Agent for Real-Time Patient Outcomes Labeling at Scale. ArXiv. 2025;abs/2509.25540.
52. Beidler P, Nguyen M, Lybarger K, Holmberg O, Ford E, Kang J. Automated Triaging and Transfer Learning of Incident Learning Safety Reports Using Large Language Representational Models. arXiv preprint arXiv:250913706. 2025.
53. Zhang QS, Kang J, Lybarger K, Glenn MC, Sponseller P, Blau KH, Ford E. Semi‐automated topic identification for radiation oncology safety event reports using natural language processing and statistical modeling. Medical physics. 2025;52(8):e17936.
54. Wang Y, De Ornelas M, Studenski MT, Bossart E, Najad-Davarani SP, Yang Y. Root Cause Analysis of Radiation Oncology Incidents Using Large Language Models. arXiv preprint arXiv:250817201. 2025.
55. Wang Y, Najad-Davarani SP, Bossart E, Studenski MT, De Ornelas M, Yang Y. Quantitative Risk Assessment in Radiation Oncology via LLM-Powered Root Cause Analysis of Incident Reports. arXiv preprint arXiv:251102223. 2025.
56. Hao Y, Holmes J, Hobson J, Bennett A, McKone EL, Ebner DK, Routman DM, Shiraishi S, Patel SH, Yu NY, Hallemeier CL, Ball BE, Waddle M, Liu W. Retrospective Comparative Analysis of Prostate Cancer In-Basket Messages: Responses From Closed-Domain Large Language Models Versus Clinical Teams. Mayo Clinic Proceedings: Digital Health. 2025;3(1):100198. doi: https://doi.org/10.1016/j.mcpdig.2025.100198.
57. Cao M, Hu S, Sharp J, Clouser EL, Holmes J, Lam LL, Ding X, Toesca DS, Lindholm WS, Patel SH, Vora SA, Wang P, Liu W. Evaluating the Performance of Using Large Language Models to Automate Summarization of CT Simulation Orders in Radiation Oncology. JACMP. 2025.
58. Verma R, Alsentzer E, Strasser Z, Chang L, Roman K, Gershanik E, Hernandez C, Linares M, Rodriguez J, Thakral D, Unlu O, You J, Zhou L, Bates D. Verifiable Summarization of Electronic Health Records Using Large Language Models to Support Chart Review. medRxiv. 2025. Epub 20250603. doi: 10.1101/2025.06.02.25328807. PubMed PMID: 40502573; PMCID: PMC12155021.
59. Small WR, Austrian J, O'Donnell L, Burk-Rafel J, Hochman KA, Goodman A, Zaretsky J, Martin J, Johnson S, Major VJ, Jones S, Henke C, Verplanke B, Osso J, Larson I, Saxena A, Mednick A, Simonis C, Han J, Kesari R, Wu X, Heery L, Desel T, Baskharoun S, Figman N, Farooq U, Shah K, Jahan N, Kim JM, Testa P, Feldman J. Evaluating Hospital Course Summarization by an Electronic Health Record–Based Large Language Model. JAMA Network Open. 2025;8(8):e2526339-e. doi: 10.1001/jamanetworkopen.2025.26339.
60. Osborne T, Abbasi S, Hong S, Sexton R, Ambut J, Patel NJ, Rosenthal RN, Ung L, Wang F, Wong R. Towards Inpatient Discharge Summary Automation via Large Language Models: A Multidimensional Evaluation with a HIPAA-Compliant Instance of GPT-4o and Clinical Expert Assessment. medRxiv. 2025:2025.04.03.25325204. doi: 10.1101/2025.04.03.25325204.
61. Holmes J, Mastroleo F, Borras-Osorio M, Seetamsetty S, Shiraishi S, Fatyga M, Boughey JC, Thiels CA, Breen WG, Ma DJ, Ebner DK, Routman DM, Laughlin BS, Vargas CE, Patel SH,

Vora SA, Laack NN, Foong AYK, Liu W, WAddle MR. The Daily Dose: Workflow-Integrated Large Language Model Automation for Clinical Summarization and Trial Identification in Radiation Oncology JAMA Network Open. 2026.
62. Jin Q, Wang Z, Floudas CS, Chen F, Gong C, Bracken-Clarke D, Xue E, Yang Y, Sun J, Lu Z. Matching patients to clinical trials with large language models. Nature Communications. 2024;15(1):9074. doi: 10.1038/s41467-024-53081-z.
63. Rybinski M, Kusa W, Karimi S, Hanbury A. Learning to match patients to clinical trials using large language models. Journal of Biomedical Informatics. 2024;159:104734. doi: https://doi.org/10.1016/j.jbi.2024.104734.
64. Shriver SP, Arafat W, Potteiger C, Butler DL, Beg MS, Hullings M, Semy S, Lister Z, Khosama L, Armstrong S, Hadley D, Pappa J, Fleury ME. Feasibility of institution-agnostic, EHR-integrated regional clinical trial matching. Cancer. 2024;130(1):60-7. Epub 20231018. doi: 10.1002/cncr.35022. PubMed PMID: 37851512.
65. Meystre SM, Heider PM, Cates A, Bastian G, Pittman T, Gentilin S, Kelechi TJ. Piloting an automated clinical trial eligibility surveillance and provider alert system based on artificial intelligence and standard data models. BMC Med Res Methodol. 2023;23(1):88. Epub 20230411. doi: 10.1186/s12874-023-01916-6. PubMed PMID: 37041475; PMCID: PMC10088225.
66. Callies A, Bodinier Q, Ravaud P, Davarpanah K. Real-world validation of a multimodal LLM-powered pipeline for high-accuracy clinical trial patient matching. Communications Medicine. 2025;5(1):536. doi: 10.1038/s43856-025-01256-0.
67. Beattie J, Owens D, Navar AM, Schmitt LG, Taing K, Neufeld S, Yang D, Chukwuma C, Gul A, Lee DS, Desai N, Moon D, Wang J, Jiang S, Dohopolski M. Large Language Model Augmented Clinical Trial Screening. medRxiv. 2024:2024.08.27.24312646. doi: 10.1101/2024.08.27.24312646.
68. Feng HY, Shan J, Vargas CE, Keole SR, Rwigema JCM, Yu NY, Ding YZ, Zhang L, Hu YL, Schild SE, Wong WW, Vora SA, Shen JJ, Liu W. Online Adaptive Proton Therapy Facilitated by Artificial Intelligence-Based Autosegmentation in Pencil Beam Scanning Proton Therapy. Int J Radiat Oncol. 2025;121(3):822-31. doi: 10.1016/j.ijrobp.2024.09.032. PubMed PMID: WOS:001434264600001.
69. Gibbons E, Hoffmann M, Westhuyzen J, Hodgson A, Chick B, Last A. Clinical evaluation of deep learning and atlas-based auto-segmentation for critical organs at risk in radiation therapy. J Med Radiat Sci. 2023;70:15-25. doi: 10.1002/jmrs.618. PubMed PMID: WOS:000857004500001.
70. Wang TY, Tam J, Chum T, Tai C, Marshall DC, Buckstein M, Liu JRY, Green S, Stewart RD, Liu T, Chao M. Evaluation of AI-based auto-contouring tools in radiotherapy: A single-institution study. J Appl Clin Med Phys. 2025;26(4). doi: 10.1002/acm2.14620. PubMed PMID: WOS:001400965300001.
71. Choi MS, Chang JS, Kim K, Kim JH, Kim TH, Kim S, Cha H, Cho O, Choi JH, Kim M, Kim J, Kim TG, Yeo SG, Chang AR, Ahn SJ, Choi J, Kang KM, Kwon J, Koo T, Kim MY, Choi SH, Jeong BK, Jang BS, Jo IY, Lee H, Kim N, Park HJ, Im JH, Lee SW, Cho Y, Lee SY, Chang JH, Chun J, Lee EM, Kim JS, Shin KH, Kim YB. Assessment of deep learning-based auto-contouring on interobserver consistency in target volume and organs-at-risk delineation for breast cancer: Implications for RTQA program in a multi-institutional study. Breast. 2024;73. doi: ARTN 103599

10.1016/j.breast.2023.103599. PubMed PMID: WOS:001126091100001.
72. Oh Y, Park S, Byun HK, Cho Y, Lee IJ, Kim JS, Ye JC. LLM-driven multimodal target volume contouring in radiation oncology. Nat Commun. 2024;15(1):9186. Epub 20241024. doi: 10.1038/s41467-024-53387-y. PubMed PMID: 39448587; PMCID: PMC11502670.

73. Wei S, Hu A, Liang Y, Yang J, Yu L, Li W, Yang B, Qiu J. Feasibility study of automatic radiotherapy treatment planning for cervical cancer using a large language model. Radiation Oncology. 2025;20(1):77.
74. Hao Y, Holmes J, Waddle MR, Davis BJ, Yu NY, Vickers KS, Preston H, Margolin D, Löckenhoff CE, Vashistha A, Kalantari S, Ghassemi M, Liu W. Personalizing prostate cancer education for patients using an EHR-Integrated LLM agent. npj Digital Medicine. 2025;8(1):770. doi: 10.1038/s41746-025-02166-0.
75. Kumarapeli P, Haddad T, de Lusignan S. Unlocking the Potential of Free Text in Electronic Health Records with Large Language Models (LLM): Enhancing Patient Safety and Consultation Interactions. Stud Health Technol Inform. 2024;316:746-50. doi: 10.3233/shti240521. PubMed PMID: 39176902.
76. Vincent P, Larochelle H, Lajoie I, Bengio Y, Manzagol P-A. Stacked Denoising Autoencoders: Learning Useful Representations in a Deep Network with a Local Denoising Criterion. J Mach Learn Res. 2010;11:3371–408.
77. AlSaad R, Abd-Alrazaq A, Boughorbel S, Ahmed A, Renault M-A, Damseh R, Sheikh J. Multimodal large language models in health care: applications, challenges, and future outlook. Journal of medical Internet research. 2024;26:e59505.
78. Truhn D, Eckardt J-N, Ferber D, Kather JN. Large language models and multimodal foundation models for precision oncology. NPJ Precision Oncology. 2024;8(1):72.
79. Sun D, Hadjiiski L, Gormley J, Chan H-P, Caoili E, Cohan R, Alva A, Bruno G, Mihalcea R, Zhou C. Outcome prediction using multi-modal information: integrating large language model-extracted clinical information and image analysis. Cancers. 2024;16(13):2402.
80. Dąbrowicki W, Rusiecki A, Jeleń Ł. Med-Agent: A Hybrid AI Agent for Multimodal Cancer Diagnosis. Procedia Computer Science. 2025;270:3290-9.
81. Truhn D, Azizi S, Zou J, Cerda-Alberich L, Mahmood F, Kather JN. Artificial intelligence agents in cancer research and oncology. Nature Reviews Cancer. 2026:1-14.
82. Assran M, Duval Q, Misra I, Bojanowski P, Vincent P, Rabbat M, LeCun Y, Ballas N, editors. Self-supervised learning from images with a joint-embedding predictive architecture. Proceedings of the IEEE/CVF conference on computer vision and pattern recognition; 2023.
83. Maes L, Lidec QL, Scieur D, LeCun Y, Balestriero R. LeWorldModel: Stable End-to-End Joint-Embedding Predictive Architecture from Pixels. arXiv preprint arXiv:260319312. 2026.